\documentstyle{elsart}

\input epsf

\begin{document}

\begin{frontmatter}

\title{Comparison of techniques to reconstruct 
VHE gamma-ray showers from multiple 
stereoscopic Cherenkov images}

\author{W.~Hofmann$^*$},
\author{I.~Jung},
\author{A.~Konopelko},
\author{H.~Krawczynski},
\author{H. Lampeitl}
\author{G.~P\"uhlhofer}

\address{Max-Planck-Institut f\"ur Kernphysik, P.O. Box 103980,
        D-69029 Heidelberg, Germany}
\address{$^*$Corresponding author (Werner.Hofmann@mpi-hd.mpg.de\\
Fax +49 6221 516 603, Phone +49 6221 516 330)}

\begin{abstract}
For air showers observed simultaneously by more than two
imaging atmospheric Cherenkov telescopes, the shower
geometry is overconstrained by the images and 
image information should be combined taking into account
the quality of the images. Different algorithms are discussed
and tested experimentally
using data obtained from observations of Mkn 501
with the HEGRA IACT system. Most of these
algorithms provide an estimate of the accuracy of the
reconstruction of shower geometry on an event-by-event basis,
allowing, e.g., to select higher-quality subsamples for 
precision measurements.
\end{abstract}

\end{frontmatter}

\section{Introduction}

The HEGRA system \cite{hegra_perf,hegra_trigger,hegra_mkn}
of imaging atmospheric Cherenkov telescopes
(IACTs) is the first installation employing the stereoscopic 
observation of air showers with multiple Cherenkov telescopes
on a routine basis. Compared to individual telescopes, IACT
stereoscopy
provides an improved reconstruction of shower
parameters and better background rejection (see, e.g., \cite{stereo}). 
Major new
instruments for VHE gamma-ray astronomy 
now in the construction phase -- such as 
VERITAS \cite{veritas} and HESS \cite{hess}
-- are based on the concept of IACT stereoscopy,
with frequently half a dozen or more telescopes observing the
same shower from different  viewing angles.

\begin{figure}[htb]
\begin{center}
\mbox{\epsfxsize10.0cm
\epsffile{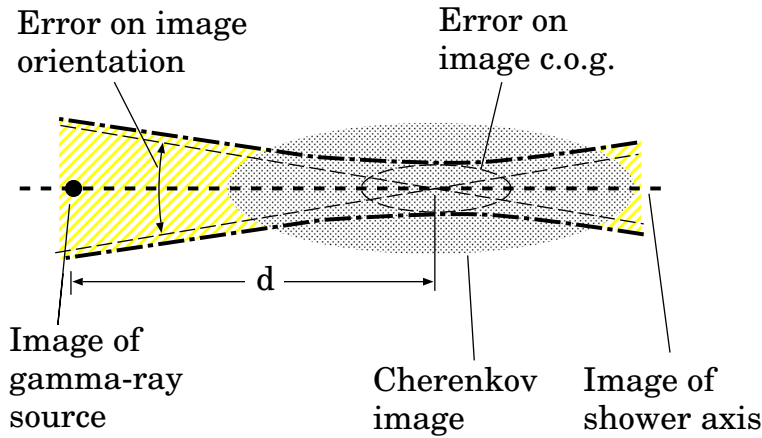}}
\caption{Cherenkov image of a gamma-ray shower
and its interpretation. The major axis
of the image approximates the image of the shower axis; the 
image of the gamma-ray source is located on the image of the
shower axis. Due to fluctuations in the shower development and
in the imaging process, the center of gravity (c.o.g.) of the
image can be displaced from the shower axis, and also the
orientation of the image can deviate. These errors are indicated
as an error ellipse for the image c.o.g., and an error on the 
image orientation. Taking into account these errors, the image of the
source is constrained to the region between the dashed-dotted lines.
Since with a simple elliptical parameterization there is a 
head-tail ambiguity of the image, the source can be located
on either side of the image. The shape of the image, in 
particular its ellipticity, can be used to estimate the shower
impact parameter relative to the telescope and hence the 
distance $d$ between the image of the source and the c.o.g.
of the Cherenkov image.}
\label{fig_image}
\end{center}
\end{figure}

In Cherenkov telescopes \cite{iact_review}, the Cherenkov light emitted by
shower particles is imaged onto a ``camera'' in the focal plane
of a large reflector, generating
an elongated, roughly elliptical image. 
The major axis of the image represents the image of the shower axis.
Therefore, the major axis of the
image points towards the image of the source on one side, and to
the point where the shower axis intersects the plane of the
telescope dish on the other side (Fig.~\ref{fig_image}). If a shower
is observed by a stereoscopic system of two Cherenkov telescopes,
its direction (i.e., the image of the source) can be determined by
superimposing the two images and intersecting their major axes. 
Similarly, the core location is obtained by intersecting
the image axes, starting from the locations of the two telescopes
(assuming that all telescope dishes are in a common plane). 
Thus, the four parameters describing the major axes of the two images
can be used to determine the four parameters describing the shower
geometry - a direction in space and an impact point in a reference plane.
It may be worth noting that the stereoscopic reconstruction of air 
showers makes the single (trivial) assumption than on average Cherenkov
images are symmetric with respect to the (image of the) shower axis.

If a shower
is observed by more than two telescopes, the shower geometry 
is overconstrained and some kind of suitable averaging or
fitting procedure is required to extract optimum shower
parameters from the information obtained from the different 
views. Particularly crucial is the case where the quality of the
information provided by the different telescopes differs
significantly, e.g. 
because one telescope is well within the light pool and sees a 
large intensity of Cherenkov light, whereas a distant telescope
may see barely enough light to provide a meaningful image. Ideally,
the reconstruction algorithm should take this difference in image
quality into account.
 This paper reviews a number of different algorithms
and describes tests of their performance based on the large
sample of gamma rays \cite{hegra_mkn} collected with the HEGRA IACT system
during the 1997 outburst of Mkn 501.

\section{Techniques to reconstruct the shower geometry}

To reconstruct shower parameters from the telescope images,
two alternative approaches can be followed:
\begin{description}
\item[Using image parameters.]
Images
provided by the different telescopes are analyzed individually, 
and their
key features are summarized in a small number of parameters (usually
the well-known Hillas parameters). Shower parameters are
derived on the basis of these image parameters.
\item[Using the full image information.]
A global optimization procure is applied to derive
the shower parameters directly from the amplitudes measured
in the individual pixels of all cameras. An example of
such techniques are global fits, where parameterized
shower images or image templates are matched to the
images observed in the different telescopes
\cite{hegra_fit,cat_fit}.
\end{description}
In this paper, we will concentrate primarily on methods
of the first type.  They are easier to
implement, and usually require significantly less
processing time. Often, analytical solutions for the
shower parameters can be derived, and one does not have
to worry about issues which arise in numerical optimization
procedures, such as the choice of proper starting values and 
the convergence
to the global optimum.
For completeness and to serve as a reference, also results based
a technique of the second type will be given.

\begin{figure}[htb]
\begin{center}
\mbox{\epsfxsize14.0cm
\epsffile{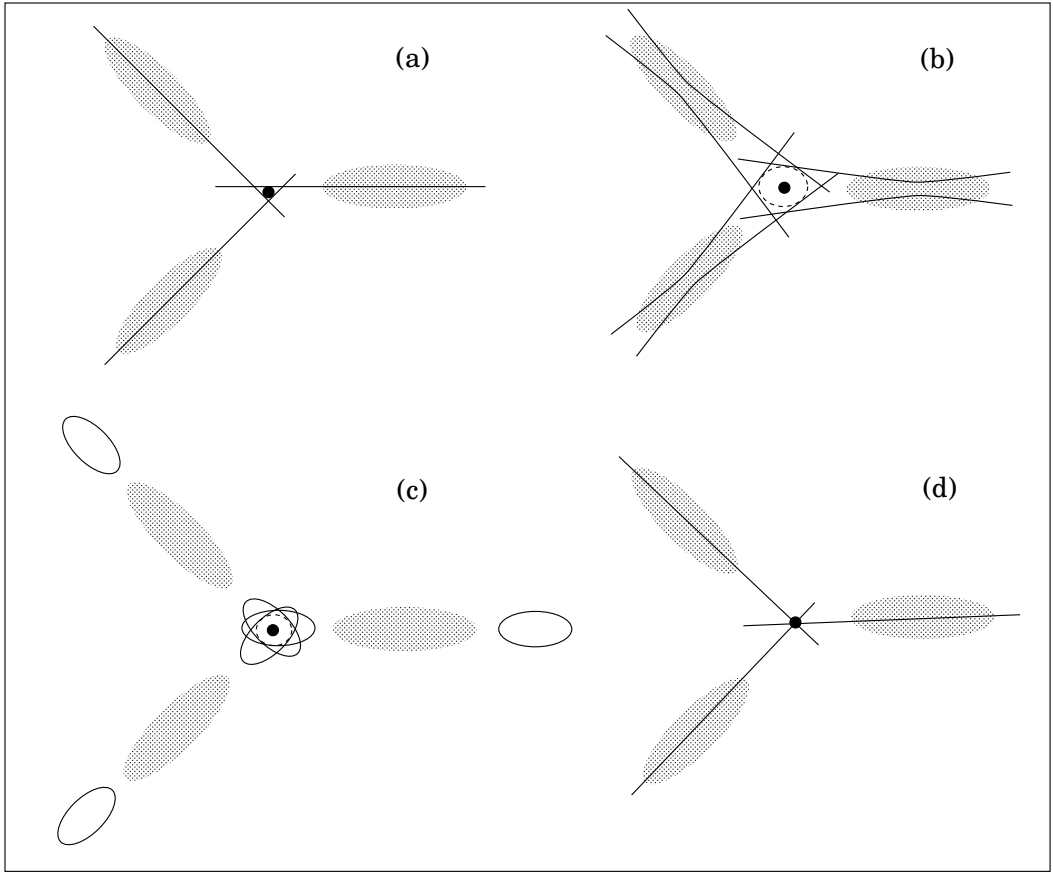}}
\caption{Illustration of different techniques to determined the
shower direction from multiple Cherenkov images. 
(a) Intersecting pairs of image axes, followed by an averaging over
intersection points. (b) Intersecting image axes taking into account
the errors on image location and image orientation, resulting in
an error ellipse for the image of the source. (c) Using in addition
the {\em width/length}-ratio to constrain the source image to two
regions on either side of an image. (d) Optimizing the shower
geometry such that the predicted image axes best match the observed
images.}
\label{fig_method}
\end{center}
\end{figure}

We will discuss seven different algorithms, six based
on the Hillas image parameters and one based on a global fit to pixel
amplitudes.
\begin{description}
\item[Algorithm 1.] For all pairs of telescopes, the image axes, derived using
the Hillas
parameterization, are intersected. In case of $N$ telescope images, the
resulting $N(N-1)/2$ intersection points are averaged, weighted with the
sine of the angle between the image axes, to take into account that image pairs
with a large stereo angle provide the most precise determination of the shower
axis. Similarly, the core location can be obtained by intersecting the image
axes, starting from the telescope locations. This technique is used, e.g., for
all published results from the HEGRA IACT system. It is illustrated in
Fig.~\ref{fig_method}(a). 
\item[Algorithm 2.] A drawback of Algorithm 1 is that differences in the
quality of the images in the different telescopes are not taken into account.
The algorithm can be improved by determining the uncertainty in the 
determination of the image c.o.g. and in the direction of the image axis,
and by taking the resulting errors 
(see Fig.~\ref{fig_image}) into account when intersecting the 
image axes, see
Fig.~\ref{fig_method}(b).
For $N$ intersecting lines with fixed error bands, the optimum
solution can be derived analytically. Since the width of the error band 
associated with each image
depends on the distance $d$ to the image c.o.g., one needs to iterate, but the
result is stable after two iterations. This method also provides
errors on the shower parameters.
\item[Algorithm 3.] The image shape contains information on the distance 
$d$ (Fig.~\ref{fig_image})
between the image c.o.g. and the image of the source. In
particular, the ratio of image {\em width} over image {\em length} can 
serve as a measure for $d$ (see also \cite{whipple_wl}). 
Smaller {\em width/length} implies large impact distance and 
large $d$.
Together with a suitable error estimate for $d$, the location,
orientation and shape of each image constrains the image of the source
 to two elliptical regions on both sides of the image (reflecting the
left-right ambiguity inherent in the parameterization of
shower images), see Fig.~\ref{fig_method}(c). For two or more images, these
error ellipses can be combined analytically to yield the optimum shower
direction and its errors. An analogous method determines the core location.
\item[Algorithm 4.] Algorithms 1,2,3 determine independently the shower
direction and the core location. Since the measurement of the image orientation
is used both in the determination of the shower direction and of the shower
core, a combined determination of core and direction should yield improved
results. Technically, for a given shower geometry, the predicted image
center lines are calculated, and a $\chi^2$ is defined measuring the agreement
of the observed image and its orientation with this prediction. Shower
geometry is chosen to minimize the sum of $\chi^2$ over telescope images.
This method is illustrated in Fig.~\ref{fig_method}(d).
\item[Algorithm 5.] Algorithm 4 can be augmented
to include the estimate of $d$ from the {\em width/length}-ratio
(see Algorithm 3), by
adding corresponding terms to the $\chi^2$.
\item[Algorithm 6.] Similar to algorithm 4, this algorithm -- proposed by
Hillas \cite{global_width}
 -- calculates the image axes for a given shower geometry, and 
varies the shower parameters such as to minimize the sum of the squared
distances of pixels to the axes, weighted with the pixel amplitudes.
This technique is analogous to the determination of the image axis for
single images, except that the weighted sum of pixel distances is minimized
for the entire set of images together, rather than for single images 
\footnote{At
first glance, Algorithm 6 may appear as an algorithm using the full pixel 
information rather than the image parameters. However, given the Hillas image
parameters, equivalent to the moments of inertia of the image with respect 
to the major and minor axis, on can easily calculate the second moment with
respect to an arbitrary axis.}.
\item[Algorithm 7.] This last algorithms makes use of the full image
information, by comparing the measured images with parameterizations of
shower images, considering the shower geometry, the energy and the 
height of the shower maximum as free parameters which are chosen to
minimize the $\chi^2$ describing the agreement between model and data.
On the basis of Monte-Carlo simulations, the technique was discussed in
\cite{hegra_fit}; a similar method was presented in \cite{cat_fit}. The
method used here differs from \cite{hegra_fit} in a different choice of
weights, which result in improved convergence.
\end{description} 
In our implementation, Algorithms 1,2 and 3 use analytic expressions
(with one iteration in case of Algorithm 3), whereas Algorithms 4
through 7 are based on numerical minimization procedures.

\section{Data sample}

To test the different algorithms and to experimentally determine the 
directional resolution achieved with each algorithm, data collected
with the HEGRA IACT system during the 1997 outburst of Mkn 501 were
used. In the HEGRA IACT system is located on the site of the 
Observatorio del Roque de las Muchachos on the Canary Island of 
La Palma at 2200 m asl. In 1997, the IACT system comprised four telescopes,
located in the center and at three sides of a square with roughly 100~m
side length. A fifth telescope at the remaining corner was integrated into
the system in 1998. The telescopes are identical, equipped with 
8.5~m$^2$ mirrors with 5~m focal length, and with 271-pixel cameras with
a diameter of the field of view of $4.3^\circ$, and an equivalent pixel
size of $0.25^\circ$. Detailed about the hardware and the data analysis
can be found in \cite{hegra_perf,hegra_trigger,hegra_mkn,hermann_padua}.

The trigger condition for individual telescopes requires a coincidence
of two pixels above a threshold of 10 photoelectrons (before June `97) or
8 photoelectrons (after June `97). 
For typical gamma-ray images, cameras trigger
once the image has more than about 40 photoelectrons.
The HEGRA IACT system as a whole is triggered and 
data are recorded whenever at least two telescopes trigger in
coincidence (see \cite{hegra_trigger} for details on the trigger system).

In the design of the HEGRA cameras and their electronics, an
important aspect was that one wanted to read out not
only those telescopes which had triggered, but also the remaining
telescopes, which will shower fainter, but frequently still usable
images.
Camera signals are digitized continuously by 120~MHz Flash-ADCs and 
are stored in a 34~$\mu$s ring buffer. A coincidence trigger
of at least two telescopes is generated with a delay of 1 to 2 $\mu$s;
After such a trigger, the readout system addresses the relevant
locations in the Flash-ADC memory and extracts the signals. 

In 1997, Mkn 501 was observed in the so-called wobble mode, with the 
source offset by $0.5^\circ$ in declination from the optical axis of
the telescopes. The offset alternated every 20~min. A region offset
by the same amount, but in the opposite direction, is used as a control
region and for background subtraction.

The analysis is based on data taken during three new-moon periods, where
the gamma-ray flux from Mkn 501 was particularly high. Only data at small
zenith angles, below $20^\circ$, are included; these showers behave 
essentially like ``ideal'' vertical showers, at least as far as the 
angular reconstruction is concerned. 
The usual selections concerning data quality were applied, see
\cite{hegra_mkn}. In total, the sample comprises
100748 events within $0.5^\circ$ from the source, and 80878 events in the 
equivalent off-source region.

Images were flat-fielded, and corrections for pointing errors of
the telescopes were applied \cite{hegra_pointing}.
Image parameters were determined by selecting ``image pixels''
as those pixels which either have a signal of 6 or more photoelectrons,
or which have a signal of at least 3 photoelectrons and are adjacent
to a pixel with 6 or more photoelectrons. 

\section{Errors assigned to image parameters, and angular resolution}

Some of the algorithms discussed above require errors on the image parameters
as input for the reconstruction of the shower axis. The relevant image
parameters are the coordinates $(x,y)$ of the center of gravity of the
images, and the direction $\theta$ of the major axis of the image. 
To parameterize the errors on the center of gravity, it is more 
convenient to use a coordinate system where axes $(u,v)$ are defined by
the major ($u$) and minor ($v$) axes of the image. In this system, the errors on
$u$, $v$, and on the orientation $\theta$ of the image should be 
essentially uncorrelated. The errors 
(in units of degr.) were parameterized on the basis of
Monte-Carlo simulations \cite{hegra_mc}:
$$
\Delta v = \left\{ {0.03 \over A} + 0.009^2 \right\}^{1 \over 2} f(w)
~~~~,~~~~~~~
f(w) = \left\{ \begin{array}{cl} 1 & \mbox{~~if~} w < 0.08 \\
                                     w/0.08 & \mbox{~~if~} w \ge 0.08
                                     \end{array} \right.
$$
and
$$
\Delta \theta = \left\{ {\left( {600 \over A}\right) }^{1.5} 
+ 1.1^2 \right\} ^{1 \over 2}
+45 \left( {w \over l} - 0.2 \right)^2
$$
Here, $A$ denotes the number of photoelectrons in the image ({\em size}),
$w$ the image {\em width} and $l$ the {\em length}. The error in
$u$ is not very relevant as long as the major axis of the image points
more or less towards the source; $\Delta u = 2 \Delta v$ was used.

As a first check to see if these errors describe the uncertainties
in the real data, the distribution in {\em miss} was plotted for 
individual telescopes of the 
(background subtracted) Mkn 501 gamma-ray sample, with the (signed)
{\em miss} parameter normalized to the expected error. The {\em miss} parameter
describes the distance between the image axis and the point on the camera
which corresponds to the image of the source. 
The error on {\em miss} is $\Delta^2_{miss} = d^2 \Delta \theta^2 + 
\Delta v^2$. For typical values of $d \approx 1^\circ$
the $\Delta \theta$ term gives the dominant contribution.
The normalized {\em miss} distribution has an rms width of 1.06, 
and its central part is well described by
a Gaussian with a width of 0.90, indicating both that the errors estimated are
accurate within 10\%, and that alignment errors of the
telescopes are small on the scale of the resolution.

\begin{figure}
\begin{center}
\mbox{\epsfxsize7.0cm
\epsffile{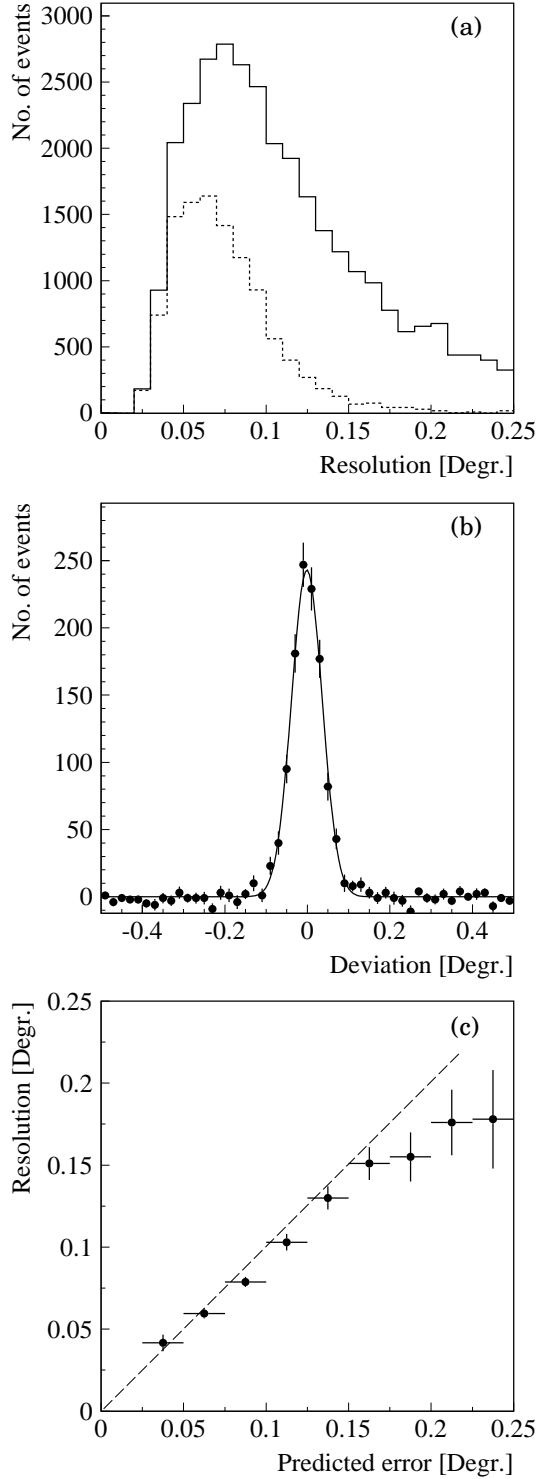}}
\caption{(a) Predicted uncertainty in the measurement of the (projected)
direction of the shower axis, for events with at least 
two triggered telescopes (full
lines) and for events where all four telescopes triggered (dashed). Only
triggered telescopes are used in the reconstruction. (b) Deviation between
the measured shower axis and the direction to Mkn 501 for events with a
predicted angular error of less than $0.04^\circ$, after statistical
subtraction of the background. The curve represents a Gaussian fit with a 
width of $0.037^\circ$. (c) Experimental angular 
resolution (in projections), determined using a 
Gaussian fit, as a function of the predicted error of the measurement.}
\label{fig_sel}
\end{center}
\end{figure}

To verify that also errors on the shower
direction can be reliably calculated by propagating the errors on the image
parameters, events were reconstructed with Algorithm 2 and
those events were selected where the predicted error on the shower
direction was less than $0.04^\circ$ (Fig.~\ref{fig_sel}(a)).
For these events, the 
distribution in the difference between the reconstructed shower direction
and the source, Mkn 501, was plotted (Fig.~\ref{fig_sel}(b)),
projected onto two orthogonal axes. Indeed, for 
this subsample of events, a (projected) angular resolution of 
$0.037^\circ$ is obtained, confirming 
the validity of the approach to estimate the errors, by treating the 
image parameters obtained by the different telescopes as independent measurements.
Here and in the following, `angular resolution' refers to the width
of the angular distribution of shower axes in a projection.
If angular resolution is defined as the half opening angle of a cone
in space, which contains 68\% of the events, the numerical values are
a factor 1.5 larger (assuming a Gaussian distribution in the errors).

Fig.~\ref{fig_sel}(c) finally shows the measured
angular resolution as a function of the predicted resolution, demonstrating
good agreement except for the tail of events with very large
predicted errors ($> 0.2^\circ$), where the measured angular
resolution is slightly better than expected
(most likely due to imperfections in the parameterizations
of the errors). Hence, already with the extremely simple
and fast Algorithm 2 one can reliably reject events with poorly reconstructed
showers, which is important, e.g., for the determination of energy spectra.

In addition to errors on the image parameters, Algorithms 3 and 5 
require an estimate of $d$ based on the image shape; we used the
empirical relations
$$
d = 1.4 - 1.25 {w \over l} ~~~~~~,~~~~~~
\Delta d = \max \left( {2.5 \over \sqrt{A}}~,~0.15 \right)
$$

\section{Comparison of reconstruction techniques}

\begin{figure}
\begin{center}
\mbox{\epsfxsize10.0cm
\epsffile{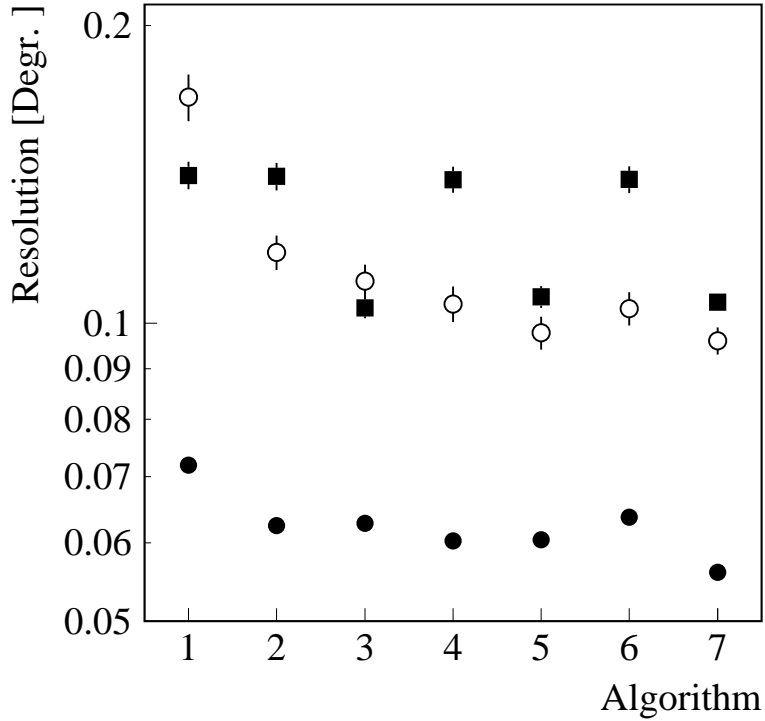}}
\caption{Angular resolution obtained from the Mkn 501 data 
with the various
algorithms, for different data samples and reconstruction modes.
Full squares: events with exactly two triggered telescopes,
using only these two telescopes; open circles: events with 
exactly two triggered telescopes, using images in all four
telescopes; full circles: events where all four telescopes
triggered and all images are used.}
\label{fig_resol}
\end{center}
\end{figure}

The angular resolutions obtained with the seven algorithms
described above are summarized in Fig.~\ref{fig_resol}, for three
characteristic data samples chosen to emphasize the specific
features on the algorithms:
\begin{description}
\item[2-Telescope events] (full squares). 
In these events, exactly two telescopes
have triggered, and only these two telescopes are used for the 
reconstruction. The 2-Telescope sample serves primarily to verify that all
algorithms work properly; unless additional shape information
is used (such as in Algorithms 3, 5, and 7), all algorithms
should give identical results if only two images are used in
the reconstruction
\footnote{It should be noted that there is a big difference between the 
2-Telescope sample,
where exactly two of the four telescopes triggered, and samples
(``2/x'')
where two telescopes are used for the reconstruction, regardless
of the state of the other two. The 2-Telescope sample selects
events which either have energies near the trigger threshold, or
which have quite distant cores. A 2/x-Telescope sample yields
for Algorithm 2
a resolution of about $0.10^\circ$, compared 
to the $0.14^\circ$ for the 2-Telescope sample -- see below.}.
\item[2+2-Telescope events] (open circles). In these events,
exactly two telescopes have triggered, but images in the other
two untriggered telescopes are included in the reconstruction.
These events represent a particular challenge to reconstruction
algorithms, since they combine images of very different quality.
Triggered images contain a mean number of about 150  
photoelectrons, compared to about 30 photoelectrons in images which did
not trigger.
\item[4-Telescope events] (full circles). In these events, all
four telescopes have triggered and are used in the reconstruction.
This class of events will obviously provide the best angular
resolution.
\end{description}

For the 2-Telescope sample (full squares in
Fig.~\ref{fig_resol}), Algorithms 1, 2, 4 and 6 do indeed
provide the identical angular resolution.
Algorithms 3, 5 and 7 -- which add shape information -- give 
significantly improved resolution. This improvement can be traced to
events with small stereo angles, i.e. with shower cores along the 
line connecting the two telescopes; for such events, the purely
geometrical reconstruction fails and the otherwise relatively
poor shape information helps to stabilize the reconstruction.

Adding now in the reconstruction the faint images
of the other two telescopes, which did not trigger -- the
2+2-Telescope sample (open circles) -- one mixes images of rather different 
quality.  If all images are combined with equal weight, as
in Algorithm 1, the faint images hurt the resolution; the
resulting resolution is
worse than if only the two triggered telescopes are used. 
In all other algorithms,
the faint images weigh less than the good images, either because
explicitly larger errors on the image parameters are assigned
(Algorithms 2-5), or because the effect of the images is weighted
with the number of photoelectrons they contain (Algorithms 6,7).
These algorithms improve the angular resolution by about 20\% to
30\% compared to the 2-Telescope sample. Little is gained by
adding the shape information via the $d(w/l)$ relation; with
four views there is always at least one reasonably large stereo angle.

In many respects, the 4-Telescope sample (full circles)
is less critical than
the 2+2-Telescope sample, since the differences between the
quality of the four images are not nearly as big. Hence, it
is no surprise that the variation between algorithms is smaller,
ranging from a resolution of $0.072^\circ$ for the worst case
(Algorithm 1) to $0.056^\circ$ for the best case (Algorithm 7).

\begin{figure}
\begin{center}
\mbox{\epsfxsize14.0cm
\epsffile{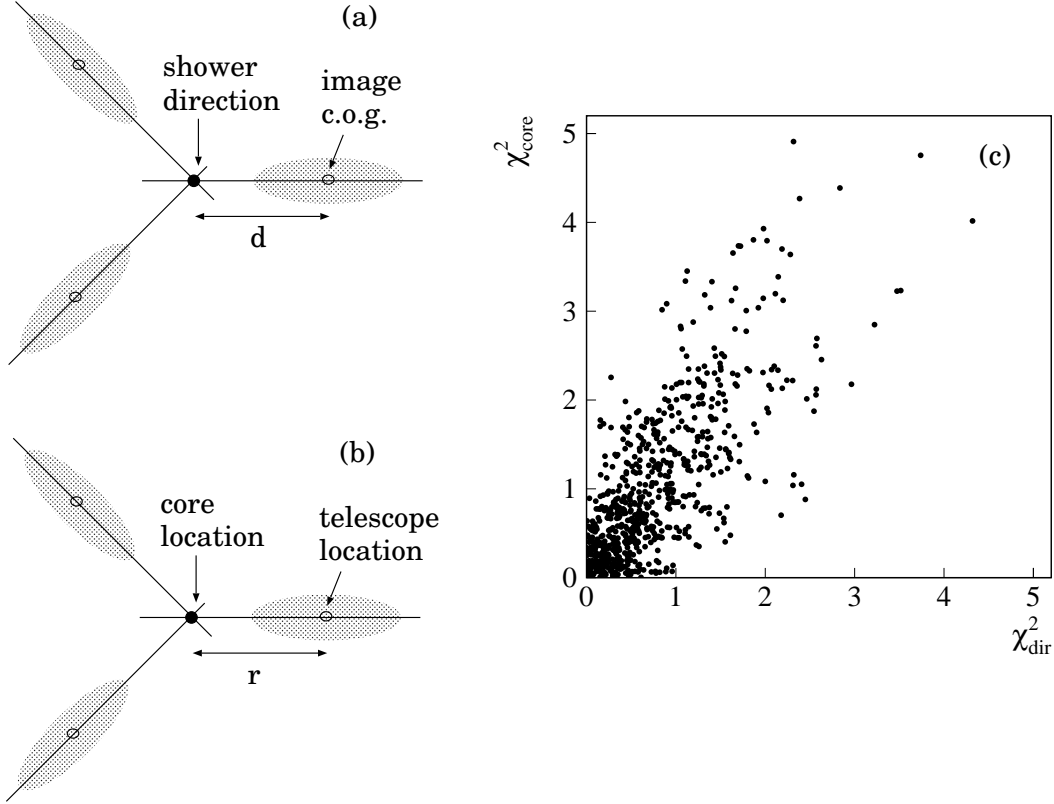}}
\caption{(a) Reconstruction of the shower direction,
by intersecting the image axes, starting from the image
c.o.g.
(b) Reconstruction of the shower core by
intersecting the image axes, starting from the telescope locations.
(c) $\chi^2$ describing the consistency in the determination 
of shower cores in overconstrained events, vs. $\chi^2$ describing
the consistency of the determination of the shower direction,
for background-subtracted gamma-ray events from Mkn 501.}
\label{fig_dircore}
\end{center}
\end{figure}

One may wonder why the joint fits of the shower direction and
of the core location (Algorithms 4, 5, 6) do not provide significant
improvements. The explanation is relatively simple, and is 
illustrated in Fig.~\ref{fig_dircore}. The geometrical figure
describing the determination of the direction (Fig.~\ref{fig_dircore}(a))
and the figure describing the determination of the core
(Fig.~\ref{fig_dircore}(b)) are essentially scaled versions of
each other, since the {\em distance} parameter $d$ of the image is
approximately proportional to the distance $r$ from the telescope
to the core location. The main difference is that in the determination
of the shower direction, the error on the position of the image
c.o.g. enters in addition to the error on the image orientation;
in the core determination, only the latter matters. Since the
error on the c.o.g. is usually of little relevance compared to the
error on the orientation, the joint fit does not add additional
constraints. Indeed, one finds that in Algorithm 2, the
$\chi^2$ describing how well the telescopes match in the determination
of the shower direction, and the $\chi^2$ of the core determination
are highly correlated (Fig.~\ref{fig_dircore}(c)).

Fitting the full image information (Algorithm 7) results in 
only very modest improvements compared to the simpler Algorithms
2 - 6. At least with the pixel size of $0.25^\circ$ of the 
HEGRA cameras, the Hillas image parameters seem to very efficiently
capture the essence of the information contained in the images.

\section{Dependence of the angular resolution on the number
of telescopes used in the reconstruction}

An interesting question is how the angular resolution depends
on the number of telescopes $N_{tel}$ used in the reconstruction.
If the 
individual images can be considered as independent, 
the resolution should improve like $1/\sqrt{N_{tel}}$.
However, at some point 
shower fluctuations will start to dominate the resolution. 

As mentioned above, to address this issue one cannot simply
use the event samples where exactly 2, 3 or 4 telescopes have
triggered, since the 2-telescope sample is biased towards
low-energy or distant showers, whereas in the 4-telescope
sample central high-energy events are enhanced.
To start from identical event samples and to avoid a ``trigger bias'',
the investigation was based on the 4-telescope sample, but only
a subset of telescopes was used to reconstruct the shower.
The resulting resolutions are illustrated in Fig.~\ref{fig_resol}(a),
for Algorithms 2 and 3. (Note that Algorithm 3 can reconstruct
the shower direction from a single image, apart from the head-tail
ambiguity.) Except for the minimum number of telescopes -- 1 for
Algorithm 3 and 2 for Algorithm 2 -- data are consistent with a  
$1/\sqrt{N_{tel}}$-dependence.

\begin{figure}
\begin{center}
\mbox{\epsfxsize8.0cm
\epsffile{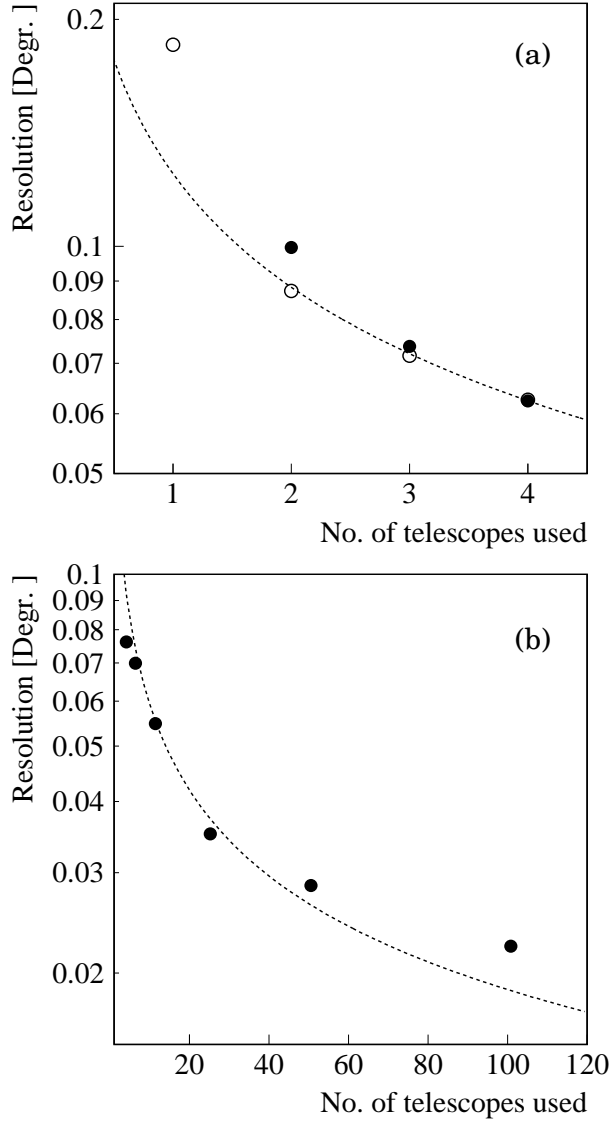}}
\caption{(a) Angular resolution obtained from the Mkn 501 data with 
Algorithm 2 (full circles) and Algorithm 3 (open
circles), for events where all four telescopes triggered,
but only a (random) subset of telescopes is used in the
reconstruction. The curve illustrates a $1/\sqrt{N_{tel}}$-dependence.
(b) Angular resolution obtained in
Monte-Carlo studies using an array of telescopes, as a 
function of the average number of telescopes used in the
reconstruction. Shower energies range from 0.5 TeV to 1 TeV.
 The curve illustrates a $1/\sqrt{N_{tel}}$-dependence.
(The telescope characteristics differ from
those of the HEGRA telescopes, and the resolutions cannot 
be compared directly.)}
\label{fig_resol1}
\end{center}
\end{figure}

The issue was further
was explored on the basis of Monte-Carlo simulations for the 
HESS telescope system \cite{hess_mc}. These simulations used
an array of 589 telescopes, arranged as a square grid of 31 x 19
telescopes, spaced 33.3~m. In the analysis, arbitrary subsets
of telescopes can be selected. The sets studied here include a set
with all telescopes turned on, a set where every other telescope
is active, and sets with telescopes on square grids with an effective 
spacing of 67~m, 100~m, 133~m, and 167~m. Only showers well
contained within the array were considered. Fig.~\ref{fig_resol1}(b)
shows the resulting angular resolution as a function of the mean number
of telescopes used in the reconstruction. 
The $1/\sqrt{N_{tel}}$-dependence of the resolution holds up to about
50 telescopes used per event,
and resolutions better than $0.03^\circ$. For even higher telescope
numbers, the dependence appears to flatten somewhat.

\section{Concluding remarks}

The main conclusions from these studies of different
algorithms for the stereoscopic reconstruction of 
multi-telescope IACT events are:
\begin{itemize}
\item By assigning and properly propagating errors of the
image parameters, reliable error estimates for the shower
direction can be obtained. It is possible to select
subsamples with improved angular resolution 
-- less than $0.05^\circ$, e.g. -- for special
purposes, such as to study the size of the source, or
to exclude poorly reconstructed events.
\item In particular when combining multiple and partly
redundant images of rather different quality, the
reconstruction algorithm must properly
account for these differences.
\item Using image shape information to constrain the
direction of the shower axis helps in the case of
2-telescope events with small stereo angles; for events
with more than two telescopes, the improvement is small.
\item Compared to the simple, robust and fast Algorithms 2 and 3,
the more `fancy' Algorithms 4, 5 and 6 as well as the 
rather sophisticated image fitting procedure of algorithm 7
give only modest improvements. For most practical purposes;
Algorithms 2 or 3 may represent the simplest and best choice.
\end{itemize}
Of course, these conclusions hold primarily for the HEGRA
Cherenkov telescopes. To which extent they can be applied to
other IACT systems depends on the degree of similarity in the
trigger concept and the layout of the cameras.

\section*{Acknowledgements}

The support of the HEGRA experiment by the German Ministry for Research 
and Technology BMBF and by the Spanish Research Council
CYCIT is acknowledged. We are grateful to the Instituto
de Astrofisica de Canarias for the use of the site and
for providing excellent working conditions. We thank the other
members of the HEGRA CT group, who participated in the construction,
installation, and operation of the telescopes. We gratefully
acknowledge the technical support staff of Heidelberg,
Kiel, Munich, and Yerevan.


\begin{thebibliography}{99}
\bibitem{hegra_perf} A. Daum et al., Astropart. Phys. 8 (1997) 1.
\bibitem{hegra_trigger} N. Bulian et al., Astropart. Phys. 8 (1998) 223.
\bibitem{hegra_mkn} F. Aharonian et al., Astron. Astrophys. 342 (1999) 69.
\bibitem{stereo} F. Aharonian et al., Astropart. Phys. 6 (1997) 343, 369.
\bibitem{veritas} Veritas Letter of Intent, T.C. Weekes et al. (1997).
\bibitem{hess} HESS Letter of Intent, F. Aharonian et al. (1997).
\bibitem{iact_review} T.C.~Weekes, Space Science Rev. 75 (1996) 1;
M. F. Cawley and T.C.Weekes, Experimental Astronomy 6 (1996) 7.
\bibitem{hegra_fit} M. Ulrich et at., J. Phys. G 24 (1998) 883.
\bibitem{cat_fit} S. Le Bohec et al., Nucl. Instr. Meth. A416 (1998) 425.
\bibitem{whipple_wl} C.W. Akerlof et al., Astroph. J. 377 (1991) L97;
J.H. Buckley et al., Astron. Astrophys. 329 (1998) 639; 
V. Connaughton et al., Astropart. Phys. 8 (1998) 179.
\bibitem{global_width} M. Hillas, talk at the Workshop on
TeV Gamma-Ray Astronomy with Systems of Cherenkov Telescopes,
Ringberg Castle (1996).
\bibitem{hermann_padua}
G.~Hermann, Proceedings of the Int. Workshop ``Towards a Major Atmospheric 
Cherenkov Detector IV'', Padua, (1995), M. Cresti (Ed.), p. 396.
\bibitem{hegra_pointing} G. P\"uhlhofer et al., Astropart. Phys. 8 (1997) 101.
\bibitem{hegra_mc} A. Konopelko et al., Astropart. Phys., in print, and
astro-ph/9901199.
\bibitem{hess_mc} A. Konopelko, Astropart. Phys., in print, and
astro-ph/9901365.
\end{thebibliography}
\end{document}